\documentclass[%
reprint,
amsmath,
amssymb, 
aps,
pra,
10pt,
]{revtex4-2}

\usepackage{graphicx}
\usepackage{dcolumn}
\usepackage{bm}
\usepackage{hyperref}
\hypersetup{
    colorlinks,
    citecolor=blue,
    filecolor=blue,
    linkcolor=blue,
    urlcolor=blue
}

\begin{document}

\preprint{APS/123-QED}

\title{Daemonic ergotropy of Gaussian quantum states and the role of measurement-induced purification via general-dyne detection}
\author{K. H. Kua}
\affiliation{Department of Physics and Astronomy, University College London,
Gower Street, WC1E 6BT London, United Kingdom}
\affiliation{ Center for Theoretical \& Computational Physics, Department of Physics, Faculty of Science, University of Malaya, Kuala Lumpur 50603, Malaysia}
\author{Alessio Serafini}
\affiliation{Department of Physics and Astronomy, University College London,
Gower Street, WC1E 6BT London, United Kingdom}
\author{Marco G. Genoni}
\affiliation{Dipartimento di Fisica “Aldo Pontremoli”, Università degli Studi di Milano, I-20133 Milan, Italy}
\date{\today}

\begin{abstract}
According to the Maxwell demon paradigm, additional work can be extracted from a classical or quantum system by exploiting information obtained through measurements on a correlated ancillary system. In the quantum setting, the maximum work extractable via unitary operations in such measurement-assisted protocols is referred to as daemonic ergotropy. In this work, we explore this concept in the context of continuous-variable quantum systems, focusing on Gaussian states and general-dyne (Gaussian) measurements. We derive a general expression for the daemonic ergotropy and examine two key scenarios: (i) bipartite Gaussian states where a general-dyne measurement is performed on one of the two parties, and (ii) open Gaussian quantum systems under continuous general-dyne monitoring of the environment. Remarkably, we show that for single-mode Gaussian states, the ergotropy depends solely on the state’s energy and purity. This enables us to express the daemonic ergotropy as a simple function of the unconditional energy and the purity of the conditional states, revealing that enhanced daemonic work extraction is directly linked to measurement-induced purification. We illustrate our findings through two paradigmatic examples: extracting daemonic work from a two-mode squeezed thermal state and from a continuously monitored optical parametric oscillator. In both case we  identify the optimal general-dyne strategies that maximize the conditional purity and, in turn, the daemonic ergotropy.
\end{abstract}
\maketitle


\section{\label{sec:intro}Introduction}
Understanding the fundamental limits imposed by quantum mechanics on work extraction from a physical system is important not only for fundamental reasons but also for practical ones. Quantum batteries, that is, quantum devices able to store and give back energy~\cite{CampaioliRMP2024}, have received a lot of attention in recent years, given also the possibility of observing a superclassical scaling in charging power~\cite{alicki2013entanglement,binderQuantacellPowerfulCharging2015,campaioliEnhancingChargingPower2017,ferraroHighPowerCollectiveCharging2018,JuliaFarreBounds2020,rossiniQuantumAdvantageCharging2020,andolina2024genuinequantumadvantagenonlinear}, but also for their relevance as energy sources for powering quantum devices such as quantum computers and quantum sensors~\cite{kurman2025,castellano2025}. First experimental attempts in these directions have been pursued by considering fluorescent molecules in optical cavities~\cite{quachSuperabsorptionOrganicMicrocavity2022,hymas2025experimentaldemonstrationscalableroomtemperature} and nuclear spins~\cite{joshiExperimentalInvestigationQuantum2022}.
From a formal point of view, ergotropy has been identified as one of the most relevant figures of merit, quantifying the maximum amount of energy extractable from a quantum state via unitary dynamics~\cite{A.E.Allahverdyan_2004}.
\begin{figure}
    \centering
    \includegraphics[width=0.95\linewidth]{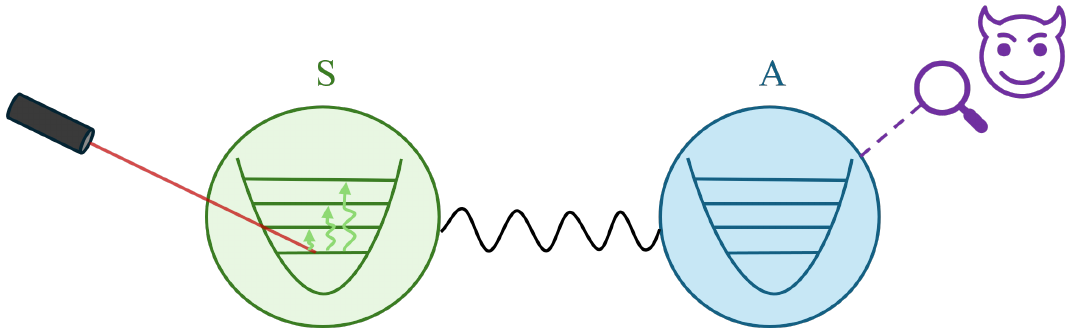}
    \caption{
    Pictorial representation of a daemonic work-extraction protocol involving two correlated quantum systems, $S$ and $A$ (here, two quantum harmonic oscillators). A quantum measurement is performed on the auxiliary system $A$, and the resulting information is exploited by a quantum Maxwell's demon to increase the extractable energy from the system $S$.
    \label{f:daemonfigure}
    }
\end{figure}

Maxwell’s demon is a thought experiment in classical thermodynamics, in which an intelligent observer reduces the entropy of a system by exploiting the information obtained by measuring the particles velocities~\cite{NoriRMPMaxwell}. Szilard later connected this idea to work extraction with his famous single-particle engine~\cite{Szilard1929}, showing that information gained through measurement can be used to extract thermodynamic work, thus linking entropy reduction and information processing.
The same approach can be taken into the quantum realm by following different approaches~\cite{junior2025friendlyguideexorcisingmaxwells,QuantumSzilardKim_PRL2011,ColletHuard2017}, and in particular the concept of daemonic ergotropy has been introduced in~\cite{francica}, as the maximum amount that can be extracted from a quantum subsystem A, by exploiting the information obtained from a measurement performed on a correlated quantum subsystem B (see Fig.~\ref{f:daemonfigure} for a pictorial representation). It is shown how the daemonic ergotropy is strictly related to the quantum correlations between the two subsystem~\cite{francica,Bernards2019}, and that in general different measurements lead to a different amount of maximum extractable work~\cite{Bernards2019,Morrone2023}.
In addition to its fundamental interest, this scenario is particularly interesting when dealing with {\em open quantum batteries}, that is, quantum batteries that undergo an open system dynamics~\cite{farina2019charger,morrone2023charging}.
In this context several studies have been done to study the impact of the environmental interaction~\cite{farina2019charger,Caravelli2021energystorage,morrone2023charging,carrega2020dissipative,Shaghaghi2023,CanzioPRA2025} and also to study the possibilities offered by such interaction~\cite{quachUsingDarkStates2020,HovhannisyanGaussErgo2020,AhmadiPRApp2025,Cavaliere2025}. However, one could also consider the physical scenario in which the environment can be continuously monitored, causing a conditional dynamics for the quantum system~\cite{Albarelli2024,WisemanMilburn}. This possibility has been extensively theoretically studied in the framework of feedback-assisted quantum state engineering~\cite{wisemanQuantumTheoryContinuous1994,dohertyFeedbackControlQuantum1999,ThomsenSpinSqueezingQuantum2002,WisemanDoherty,genoniQuantumCoolingSqueezing2015,brunelliConditionalDynamicsOptomechanical2019,digiovanniUnconditionalMechanicalSqueezing2021,candeloroFeedbackAssistedQuantumSearch2023,isaksenMechanicalCoolingSqueezing2023,Caprotti_2024} quantum estimation~\cite{mabuchiDynamicalIdentificationOpen1996,geremiaQuantumKalmanFiltering2003,tsangOptimalWaveformEstimation2009,gammelmarkFisherInformationQuantum2014,sixParameterEstimationMeasurements2015,kiilerichBayesianParameterEstimation2016,genoniCramErRaoBound2017,albarelliUltimateLimitsQuantum2017,Albarelli2018restoringheisenberg,rossiNoisyQuantumMetrology2020,AmorosBinefa2021,FallaniPRXQuantum2022,ilias2022criticality,yangEfficientInformationRetrieval2022,amorosbinefa2024,midha2025metrologyopenquantumsystems,khan2025tensornetworkapproachsensing,yang2025quantumcramerraoprecisionlimit} and out-of-equilibrium quantum thermodynamics~\cite{manzanoQuantumThermodynamicsContinuous2022,garrahanThermodynamicsQuantumJump2010,rossiExperimentalAssessmentEntropy2020,landiInformationalSteadyStates2022}, with experimental demonstration of such protocols ranging from superconducting circuits~\cite{murchObservingSingleQuantum2013,campagne-ibarcqObservingQuantumState2016,ficheuxDynamicsQubitSimultaneously2018,minevCatchReverseQuantum2019} to optomechanics~\cite{wieczorekOptimalStateEstimation2015,rossiObservingVerifyingQuantum2019,rossiMeasurementbasedQuantumControl2018,magriniRealtimeOptimalQuantum2021} and atomic systems~\cite{duan_concurrent_2025}. If we are interested in enhancing energy storage and extraction, the information obtained from the environment continuous monitoring can in fact be exploited in two ways: either in the charging protocol, with the aim of preparing quantum battery states with high ergotropy via feedback~\cite{mitchisonChargingQuantumBattery2021,yaoOptimalChargingOpen2022a} or, following the Maxwell demon paradigm described above, in the work extraction phase. The concept of {\em unravelling daemonic ergotropy} has then been introduced to address this specific instance, quantifying the maximum work that can be extracted from an open quantum battery whose interacting environment is continuously monitored in time~\cite{Morrone2023,Elyasi2025}. 

Experimental demonstrations of such daemonic work extraction protocols have recently been shown, for example, by simulating a continuously monitored quantum battery via a collisional model on a digital quantum computer~\cite{Elyasi2025}, and by considering measurements on correlated trapped ions~\cite{stahl2024demonstrationenergyextractiongain}.\\

Continuous-variable quantum systems describe different physical experimental scenarios, ranging from quantum optical platforms, quantum optomechanical, and atomic systems, and they have been extensively studied as basis for quantum technologies such as quantum computation, quantum communication, and quantum metrology~\cite{SERAFINI_2023}. Here, we focus on Gaussian dynamics, such that the corresponding quantum states are fully described by first and second moments of the quadrature operators (i.e. position and momentum operators)~\cite{SERAFINI_2023,Diffusone,GaussFOP}. 
Starting from the general formula for the ergotropy of Gaussian states, already presented in the literature~\cite{HovhannisyanGaussErgo2020,rodriguez2025extractingenergybosonicgaussian}, we derive the corresponding one for the daemonic ergotropy under general-dyne measurements. We discuss both the scenario of bipartite Gaussian states where one of the parties is measured and work is extracted from the other party via a (conditional) unitary operation, and the open quantum system (dynamical) scenario, where the environment is continuously monitored via general-dyne detection; in the latter case, the daemonic ergotropy quantifies the maximum average work extractable by performing a (conditional) unitary on each quantum trajectory. 

In particular, we focus on single-mode Gaussian state: we unveil that the ergotropy is in fact a simple function of the state energy and purity; as a consequence we prove that the daemonic ergotropy is a simple function of the unconditional state energy and of the conditional state purity, which for Gaussian dynamics is deterministic, that is does not depend on the measurement outcome, but only on the kind of general-dyne monitoring performed. We thus find that optimizing the measurement-enhanced extractable work is in fact completely equivalent to optimize the conditional state purity, and thus to optimize a {\em cooling} protocol (from now onward we will use in an equivalent way the terms cooling and purification). While purification under general-dyne monitoring has already been addressed for discrete variable systems~\cite{jacobsHowProjectQubits2003,combesRapidStateReduction2006,wisemanReconsideringRapidQubit2006,jordanQubitFeedbackControl2006,wisemanOptimalityFeedbackControl2008,ruskovQubitPurificationSpeedup2012}, such analysis has never been performed for continuous-variable (Gaussian) systems. We then study in detail the optimization of daemonic ergotropy and conditional purity when possible in general, and then by discussing two paradigmatic examples : i) a two-mode squeezed thermal states under partial general-dyne detection, where we show that heterodyne measurement is the optimal strategy; ii) a continuously monitored optical parametric oscillator. In this second example we also observe that the optimal general-dyne strongly depends on the physical parameters characterizing the system and that, somehow unexpectedly, the unravelling leading to larger purification and daemonic ergotropy does not correspond to the one leading to larger values of conditional quantum squeezing. \\

The manuscript is structured as follows: in Sec.~\ref{s:Gaussian} we give a brief but comprehensive introduction to the Gaussian formalism for continuous-variable quantum systems, including the case of continuously monitored open scenario; in Sec.~\ref{s:ergotropy} we discuss work extraction from quantum systems, introducing the concept of ergotropy and of daemonic ergotropy; in Sec.~\ref{s:ergo_gauss} we recall the general formula for the ergotropy of Gaussian states, showing in particular how this simplifies for single-mode states. In Sec.~\ref{s:dergo_gaussian_measured} we present the formula of the daemonic ergotropy given a bipartite Gaussian state, and we present an analytical formula for the maximum daemonic ergotropy both for the optimal general-dyne measurement and also for the (experimental relevant) sub-cases of optimal homodyne and heterodyne, by finally focusing on the special case of two-mode squeezed thermal states. In Sec.~\ref{s:dergo_gaussian_open} we finally discuss the unravelling daemonic ergotropy for open Gaussian system under general-dyne monitoring, providing some general results for the steady-state purification under general-dyne monitoring, and by then discussing in detail the case of a continuously monitored optical parametric oscillator. We conclude the manuscript in Sec.~\ref{s:conclusions} with some concluding remarks and outlooks.
\section{Gaussian quantum states formalism}\label{s:Gaussian}
\subsection{Gaussian states and Gaussian unitaries}
The $2n$ canonical operators of a bosonic system of $n$ modes can be arranged in a vector of operators 
$    \mathbf{\hat{r}}=(\hat{x}_1,\hat{p}_1,...,\hat{x}_n\hat{p}_n)^\top$, 
obeying the canonical commutation relations 
$
[\mathbf{\hat{r}},\mathbf{\hat{r}^\top}]=\mathbf{\hat{r}}\mathbf{\hat{r}^\top} - (\mathbf{\hat{r}}\mathbf{\hat{r}^\top})^\top = i\Omega$, where natural units and the outer product notation have been adopted \cite{SERAFINI_2023}, and the antisymmetric symplectic form $\Omega$ is given by
\begin{equation} \label{eqn:symplecticform}
        \Omega=\bigoplus^n_{j=1}\Omega_1, \indent  \text{with} \indent \Omega_1=  \begin{pmatrix}
0 & 1 \\
-1 & 0 
\end{pmatrix}.  
\end{equation}
A general second-order Hamiltonian can be written as
$\hat{H}=\frac{1}{2}\mathbf{\hat{r}^\top}H\mathbf{\hat{r}} + \mathbf{\hat{r}^\top}\mathbf{r}$,
where $H$ is a real, symmetric matrix of Hamiltonian couplings and $\mathbf{r}$ is a vector of linear driving terms.
Equivalently, up to an additive constant, one may write
$\hat{H} = \frac{1}{2}\hat{D}_{-\mathbf{\Bar{r}}}\mathbf{\hat{r}}^\top H \mathbf{\hat{r}}\hat{D}_{\mathbf{\Bar{r}}}$, 
where $\mathbf{\Bar{r}}= -H^{-1} \mathbf{r}$ and
the Weyl (displacement) operators $ \hat{D}_{\mathbf{\Bar{r}}}$ are defined as  
$ \hat{D}_{\mathbf{\Bar{r}}} =e^{i\mathbf{\Bar{r}^\top}\Omega\mathbf{\hat{r}}} 
$ and act as translations on the canonical operators:
$\hat{D}_{\mathbf{\Bar{r}}}\mathbf{\hat{r}}\hat{D}^{\dag}_{\mathbf{\Bar{r}}}=\mathbf{\hat{r}}+\mathbf{\Bar{r}}$.

Then, the set of Gaussian states may then be defined as the ground and thermal states of second-order Hamiltonians with a positive definite Hamiltonian matrix $H>0$ \cite{SERAFINI_2023}:
\begin{equation} \label{eqn:generalgaussstate}
    \rho_G=\frac{e^{-\beta\hat{H}}}{\text{Tr}[e^{-\beta\hat{H}}]},
\end{equation}
where $\beta \in \mathbb{R}^+$ and $\hat{H}$ is a second-order Hamiltonian, as defined above (this definition is equivalent to assuming a Gaussian Wigner function).
Alternately, Gaussian states are entirely characterised by the vector of first moments and the covariance matrix (CM):
\begin{equation} \label{eqn:gauspara}
\begin{split}
        \mathbf{\Bar{r}}&=\text{Tr}[\rho_G \mathbf{\hat{r}}] \\ \bm\sigma&=\text{Tr}[\{ (\mathbf{\hat{r}}-\mathbf{\Bar{r}}),(\mathbf{\hat{r}}-\mathbf{\Bar{r}})^\top \}\rho_G].
\end{split}
\end{equation}
The first moments can take any real values, while the CM is constrained by the Heisenberg uncertainty relation  $\bm\sigma+i\Omega\ge 0$.

Unitaries generated by purely quadratic Hamiltonians form an infinite dimensional (metaplectic) representation of the real symplectic group $Sp_{ 2_n,\mathbb{R}}$. The latter is defined as the set of real matrices of dimension $2n\times2n$ that preserve $\Omega$, the symplectic form, under congruence: $S\in Sp_{ 2_n,\mathbb{R}} \indent \iff \indent S \Omega S^\top=\Omega$.
More specifically, a symplectic transformation $S_H={\rm e}^{\Omega H}$ represents the Heisenberg evolution of the vector of canonical operators $\mathbf{\hat{r}}$ under the quadratic Hamiltonian $\hat{H}$ defined above, where $\mathbf{r}=\mathbf{\Bar{r}}=0$ (that is what is meant by ``purely quadratic'').

Gaussian states described by quadratic Hamiltonians can always be diagonalised using displacements and symplectic transformations \cite{SERAFINI_2023}. 
This is achieved by the displacement that sets the first moments to zero, followed by the symplectic transformation which puts the CM $\bm\sigma$ in normal form (analogous to the normal mode decomposition of a quadratic Hamiltonian, which always exists since $\bm\sigma>0$):
\begin{equation} \label{eqn1:covariancematrixofgaus}
    \bm\sigma = S \left(\bigoplus_{j=1}^n \nu_j\openone_2\right) S^\top,
\end{equation}
where the $\nu_j$'s are known as the symplectic eigenvalues of $\bm\sigma$, corresponding to the (at least twice degenerate) eigenvalues of $|\Omega\bm\sigma|$. 
The spectrum of a Gaussian state is therefore a function of its symplectic eigenvalues alone, which we report here since it will have a major bearing on our discussion about ergotropy:

Note also that the normal mode decomposition of a CM corresponds to rotating the Gaussian state into a thermal state of the ``free Hamiltonian'' 
\begin{align}
\hat{H}_0 = \frac12\sum_{j}(\hat{x}_j^2+\hat{p}_j^2).
\label{eq:freehamiltonian}
\end{align}
In terms of the density operator this corresponds to the formula
\begin{align}
\rho_G = \hat{D}_{\bar{\bf r}} \hat{U}_S \left( \bigotimes_{j=1}^n \hat{\nu}_j \right)  \hat{U}_S^\dag \hat{D}_{\bar{\bf r}}^\dag ,  
\label{eq:rhoGauss}
\end{align}
where $\hat{U}_S$ denotes the Gaussian unitary evolution corresponding to the symplectic matrix $S$ and $\hat{\nu}_j$ are the aforementioned thermal states of the single-mode free Hamiltonians $\hat{H}_{0,j} = \frac{\omega_j}2 (\hat{x}_j^2 + \hat{p}_j)^2$. Notice that the thermal average excitations $n_j$ of these thermal states are related to the symplectic eigenavlues of the CM $\nu_j$ via the formula $\nu_j = (2 n_j +1) = 1 +2 (e^{\beta\omega_j} -1)^{-1}$.
As stated above, all the properties of a Gaussian quantum state can be evaluated from its first moment vector $\bar{\bf r}$ and CM $\bm\sigma$; in particular one can easily show that its energy (in terms of the free Hamiltonian) reads
\begin{align}
    E(\rho_G) = \hbox{Tr}[\rho_G \hat{H}_0] = \frac12 |\Bar{\bf r}|^2 + \frac14 {\rm tr}[\bm\sigma] \,,
    \label{eq:energy}
\end{align}
where ${\rm Tr}[\cdot]$ and ${\rm tr}[\cdot]$ denote respectively the trace of operators in the Hilbert space and of finite dimensional matrices. Similarly one can prove that the purity of a Gaussian quantum state is simply equal to
\begin{align}
    \mu(\rho_G) = \hbox{Tr}[\rho_G^2] =\frac{1}{\sqrt{{\rm det}[\bm\sigma]}} \,.
    \label{eq:purity}
\end{align}

As we saw, Gaussian unitaries can be decomposed into linear operations (displacement operators) and purely quadratic operations (symplectic transformations). The former just shift the first moments by an arbitrary amount (requiring, as one should expect, an arbitrary amount of energy). The latter admit a useful singular-value -- also occasionally referred to as ``Euler'' or ``Bloch-Messiah'' -- decomposition (SVD)\cite{SERAFINI_2023}:
\begin{equation} \label{eqn:svd}
    S=O_1 Z O_2,
\end{equation}
with $O_1, O_2 \in S_{P2_n,\mathbb{R}} \cap SO(2n)$ which is the compact subgroup of orthogonal symplectic matrices and
\begin{equation}
    Z=\bigoplus^n_{j=1} \begin{pmatrix}
z_j & 0 \\
0 & z_j^{-1} 
\end{pmatrix}, \quad z_j>0 \quad \forall j\in [1,....,n].
\end{equation}
Symplectic transformations are thus split into passive ($O_1$ and $O_2$) and active, ``squeezing'' ($Z$) transformations, a terminology which refers to the fact that the passive orthogonal transformations, implemented in optics through phase shifters and beam splitters, commute with the free Hamiltonian, whilst squeezers do not.
\subsection{General-dyne measurements and conditional Gaussian dynamics}
There exists a class of measurements -- or, more technically, of positive operator valued measures (POVMs) -- that preserve the Gaussian character of the measured state and are readily available in practice. Such measurements are also often dubbed ``Gaussian'' but we shall stick to the more exact denomination of ``general-dyne POVMs'', defined by the following resolutions of the identity \cite{SERAFINI_2023}:
\begin{equation}\label{povm}
    \openone = \frac{1}{(2\pi)^n} \int_{\mathbb{R}^{2n}} d\mathbf{r}_m \hat{D}_{-\mathbf{r}_m} \rho_m \hat{D}_{\mathbf{r}_m}.
\end{equation}

\noindent Here, $\rho_m$ is a Gaussian state with null first first moments and second moments $\bm\sigma_{\sf m}$ obeying the Heisenberg relation $\bm\sigma_{\sf m} + i\Omega \geq 0$. The $2n$ real displacement parameters $\mathbf{r}_m$ label the outcome of the general-dyne measurement process, which is characterised by $\bm\sigma_{\sf m}$. The outcomes $\mathbf{r}_m$ occur, given a measured state $\rho$, with probability density \cite{SERAFINI_2023}

\begin{equation}
    p(\mathbf{r}_m) = \frac{\text{Tr}[\rho \hat{D}_{-\mathbf{r}_m} \rho_m \hat{D}_{\mathbf{r}_m}]}{(2\pi)^n}.
\end{equation}
If the system one is measuring is also a Gaussian state, $\rho$ with first and second moments, $\mathbf{\Bar{r}}$ and $\bm\sigma$ respectively, the probability density becomes Gaussian:

\begin{equation} \label{pgau}
    p(\mathbf{r}_m) = \frac{e^{-(\mathbf{r}_m - \mathbf{\Bar{r}})^\top(\bm\sigma + \bm\sigma_{\sf m})^{-1}(\mathbf{r}_m-\mathbf{\Bar{r}})}}{\pi^n \sqrt{\text{Det}(\bm\sigma + \bm\sigma_{\sf m})}}.
\end{equation}
Notice the expediency of this parametrisation, where a $\bm\sigma_{\sf m}$ corresponding to a mixed quantum state, i.e. such that $\det\bm\sigma_{\sf m}\ge 1$, represents noisy measurements. 

The case $\bm\sigma_{\sf m}=\openone$ corresponds to ``heterodyne detection'' (projection on the non-orthogonal but complete set of coherent states) while, in the single-mode case and up to phase space rotations, the limit $\bm\sigma_{\sf m}=\lim_{z_{\sf m}\rightarrow\infty}{\rm diag}(z_{\sf m},1/z_{\sf m})$ (in which case the measurement outcome reduces to a single real value) represents the instance of ``homodyne detection''. 

In general, by applying the normal-mode decomposition (\ref{eqn1:covariancematrixofgaus}) and the SVD (\ref{eqn:svd}), one may parametrise single-mode general-dyne detections with three parameters $\nu_m$, $\vartheta_{\sf m}$ and $z_{\sf m}$. These determine, respectively, imperfections, the optical phase and the type of general-dyne projection, such that $\bm\sigma_{\sf m}=\nu SS^{\sf T}$ and $S=R_{\vartheta_{\sf m}}{\rm diag}(\sqrt{z_{\sf m}},1/\sqrt{z_{\sf m}})$, where $R_{\vartheta_{\sf m}} $ is a two-dimensional rotation by the angle $R_{\vartheta_{\sf m}}$. We remark that, in the quantum optics scenario, any general-dyne detection can be implemented via the standard double-homodyne scheme designed to implement heterodyne detection, and by tuning a beam-splitter transmissivity according to the desired value of $z_m$~\cite{GeneralDino}.

With regard to the parameter $\nu_m=\sqrt{{\rm Det}\bm\sigma_{\sf m}}$, note that the POVM (\ref{povm}) may also be written as 
\begin{equation}\label{eqn:cmpovm}
    \openone = \frac{1}{2\pi^2} \int_{\mathbb{R}^{2}}\hspace*{-0.2cm} {\rm d}\mathbf{r}_m \int_{\mathbb{R}^{2}} \hspace*{-0.2cm}{\rm d}\mathbf{y} {\rm e}^{-\mathbf{y}^{\sf T}Y^{-1} \mathbf{y}} \hat{D}_{\mathbf{r}_m-\mathbf{y}}^\dag |\psi_m\rangle\langle\psi_m| \hat{D}_{\mathbf{r}_m-\mathbf{y}},
\end{equation}
where $Y=(\nu_m-1)SS^{\sf T}\ge0$ and $|\psi_m\rangle$ is the pure Gaussian state with CM $SS^{\sf T}$. This relationship holds because any noisy CM can be derived from a pure one by adding noise through normally distributed displacements — a CP-map known as ``classical mixing’’ \cite{SERAFINI_2023}. Hence, {\em any noisy general-dyne measurement is equivalent to a convex combination of ideal, pure, displaced general-dyne projections}. This will be key to maximising the average ergotropy with respect to $\nu$.

If one now considers a multipartite Gaussian state, partitioned into subsystems $A$ and $B$, as

\begin{equation}
    \bm\sigma = \begin{pmatrix}
        \bm\sigma_A & \bm\sigma_{AB} \\ \bm\sigma_{AB}^\top & \bm\sigma_B
    \end{pmatrix} \quad \text{and} \quad \mathbf{\Bar{r}} = \begin{pmatrix}
        \mathbf{\Bar{r}}_A \\ \mathbf{\Bar{r}}_B,
    \end{pmatrix} \label{eq:2ms}
\end{equation}

\noindent and has subsystem $B$ subjected to general-dyne measurements, the final {\em conditional} state of subsystem $A$ (i.e., the state filtered upon the knowledge of outcome $\mathbf{\Bar{r}}_B$) is a Gaussian state whose statistical moments are mapped according to \cite{SERAFINI_2023}

\begin{equation}\label{eqn:gendynemapCM}
    \bm\sigma_A \mapsto \bm\sigma_A^{(c)}= \bm\sigma_A - \bm\sigma_{AB}(\bm\sigma_B+\bm\sigma_{\sf m})^{-1}\bm\sigma_{AB}^\top,
\end{equation}
\begin{equation} \label{eqn:gendymapR}
    \mathbf{\Bar{r}}_A \mapsto \mathbf{\Bar{r}}_A^{(c)} =  \mathbf{\Bar{r}}_A + \bm\sigma_{AB}(\bm\sigma_B+\bm\sigma_{\sf m})^{-1}(\mathbf{r}_m-\mathbf{\Bar{r}}_B).
\end{equation}


\subsection{Continuously monitored Gaussian quantum systems}
\label{s:gaussian_open}
The open, diffusive dynamics of a Gaussian state interacting bi-linearly with an environment in the Born-Markov regime may be dealt with through the input-output formalism \cite{Gardiner_1985,Diffusone,Albarelli2024}.
The input is represented as a $2m$-dimensional vector of operators, $\mathbf{\hat{r}}_{in}(t)$ (where the label $t$ implies interaction with the system at a time $t$ for an infinitesimal interval $dt$), that interact with the system of $n$ modes through the general coupling Hamiltonian 
\begin{equation} \label{eqn1: arbitrarycoupling}
    \hat{H}_C = \mathbf{\hat{r}}^\top C \mathbf{\hat{r}}_{in} 
\end{equation}
where 
$C$ is a completely  arbitrary $2n\times 2m$ real coupling matrix. 
The environmental output modes following the interaction, $\mathbf{\hat{r}}_{out}(t)$, may be in principle -- and oftentimes in practice -- be monitored, as we shall assume later on. 


If the input modes satisfy the white noise condition $[\mathbf{\hat{r}}_{in}(t),\mathbf{\hat{r}}_{in}(t)^{\sf T}]=i \Omega \delta(s-t)$, proper to a completely memory-less bath, then, by introducing the standard It\^{o} rule, applying the symplectic formalism and tracing out the environment, one obtains the following {\em unconditional} diffusive dynamics for the first and second statistical moments \cite{SERAFINI_2023} (i.e., the dynamics if monitoring of the environment does not take place, or is forgotten): 

\begin{equation} \label{eqn1: unconditionalfirst}
    \frac{d\mathbf{\Bar{r}}_{\sf unc}}{dt} = A\mathbf{\Bar{r}}_{\sf unc} + \mathbf{d},
\end{equation}

\begin{equation} \label{eqn1 unconditional dynamics second}
    \frac{d\bm\sigma_{\sf unc}}{dt}=  A\bm\sigma_{\sf unc} + \bm\sigma_{\sf unc} A^\top + D,
\end{equation}

\noindent with

\begin{equation} \label{eqn1: driftdiffu}
\begin{split}
    A &= \Omega H_S + \frac{\Omega C \Omega C^\top}{2} \\ D&=\Omega C \bm\sigma_{\sf in} C^\top \Omega^\top\\ \mathbf{d} &= \Omega C \mathbf{\Bar{r}}_{in} ,
\end{split}
\end{equation}
where $\bm\sigma_{\sf in}$ and $\mathbf{\Bar{r}}_{in}$ are the second and first moments of the input fields (with the latter giving rise to the finite linear driving term $\mathbf{d}$).

If we now assume that the output modes $\mathbf{\hat{r}}_{out}(t)$ are continuously measured via a general-dyne measurement $\bm\sigma_{\sf m}$, by applying the mappings (\ref{eqn:gendynemapCM}) and (\ref{eqn:gendymapR}) one can derive the modified equations of motions of a Gaussian state, conditioned on the output of the measured photocurrent, which reads~\cite{SERAFINI_2023,Diffusone}
\begin{align}
d{\bf y} = - B^T \bar{\bf r}_c \,dt + d{\bf w}_t \,,
\end{align}
where $B=C\Omega(\bm\sigma_{\sf in}+\bm\sigma_{\sf m})^{-1/2}$. The vector $d\mathbf{w}_t$ operationally corresponds to the {\em innovation} term in the photocurrent (that is the difference between the actual result of the measurement and its expected average value) and mathematically is proven to be a vector of independent Wiener increments,  satisfying the statistics (we denote with $\mathbb{E}[\cdot]$ the average over all the possible measurement results/trajectories):

\begin{equation} \label{eqn2: wienerstatistics}
    \mathbb{E}[d\mathbf{w}_t ] =0 \quad \text{and} \quad  \mathbb{E}[ \{d\mathbf{w}_t,d\mathbf{w}_t^\top \}] = \openone dt\,.
\end{equation}
In fact the evolution of the first moments under monitoring is indeed a stochastic process, depending on the innovation term $d\mathbf{w}_t$ as follows
\begin{equation}\label{eqn:gendynemonifirst}
      d\mathbf{\Bar{r}}_c = A \mathbf{\Bar{r}}_c dt + \mathbf{d} dt + (E-\bm\sigma B) d\mathbf{w}_t.
\end{equation}
On the other hand, for the second moments, one obtains the deterministic Riccati equation (the independence of the conditional second-moments from these monitoring outcomes is a highly non-trivial feature of Gaussian dynamics):
\begin{equation} 
    \frac{d\bm\sigma_c}{dt} = {A} \bm\sigma_c + \bm\sigma_c {A}^\top+ {D}- (E - \bm\sigma_c B)(E - \bm\sigma_c B)^\top
    \label{eq:riccati}
\end{equation}
with $E=\Omega C \bm\sigma_{\sf in}(\bm\sigma_{\sf in}+\bm\sigma_{\sf m})^{-1/2}$.

We remark that thanks to the linearity of the trace, one has that $\mathbf{\Bar{r}}_{\sf unc} = \mathbb{E}[\mathbf{\Bar{r}}_c]$, while for the CM one has $\bm\sigma_{\sf unc}=\bm\sigma_c + \bm\Sigma$, where the excess noise matrix $\bm\Sigma=\mathbb{E}[\{ \mathbf{\Bar{r}}_c,\mathbf{\Bar{r}}_c\}] - \{\mathbb{E}[\mathbf{\Bar{r}}_c],\mathbb{E}[\mathbf{\Bar{r}}_c]\}$ quantifies the first moments stochasticity.\\

The formalism above, in terms of first moment vector and CM, is clearly equivalent to the one in the Hilbert space, where the dynamics of the corresponding conditional quantum state $\rho_c$ is described by a diffusive stochastic master equation, and the dynamics of the unconditional quantum state $\rho_{\sf unc}=\mathbb{E}[\rho_c]$ is described by a Markovian master equation in the Lindblad form~\cite{Albarelli2024}.
\section{Ergotropy and daemonic ergotropy}
\label{s:ergotropy}
Ergotropy is defined as the maximum amount of work that can be extracted from a quantum state $\varrho$ through a unitary cyclic process, keeping its entropy constant. In formulae, by denoting with $\hat{H}_0$ the Hamiltonian that defines the energy we want to extract, this corresponds to the quantity~\cite{A.E.Allahverdyan_2004}

\begin{equation} \label{eqn1: generalergo}
    \mathcal{E}(\rho)= {\rm Tr}\left[\rho \hat{H}_0\right] -
    \min_{\hat{U}}{\rm Tr}\left[\hat{U} \rho \hat{U}^\dag\hat{H}_0 \right] , 
\end{equation}
where the minimisation runs over all unitary operations in the system dimension. 
Notice that such a minimisation is attained by the unitary that maps all the state eigenvectors to energy eigenvectors, such that the one corresponding to the largest eigenvalue is assigned to the lowest energy and all others to higher energy levels in decreasing order. Such states are called {\em passive}, as they do not allow for the reversible extraction of work. One can thus write equivalently the ergotropy as
\begin{equation} \label{eq:ergo_passive}
    \mathcal{E}(\rho) = {\rm Tr}\left[\rho \hat{H}_0\right] -{\rm Tr}\left[\tau_\rho\hat{H}_0 \right] , 
\end{equation}
where we have denoted with $\tau_\rho$ the passive state corresponding to the initial state $\rho$. One should notice that thermal states of a certain Hamiltonian $\hat{H}_0$ are passive with respect to it, in view of their exponentially decreasing Boltzmann occupation of the energy levels, but in general thermal states do not correspond to the whole set of passive states. As described above, one can readily identify the passive state $\tau_\rho$, once the eigenvalues and eigenvectors of  $\rho$ and $\hat{H}_0$ are given. This allows to both identify the optimal unitary $\hat{U}$ yielding the maximum extractable work, and to write a compact closed formula for the ergotropy~\cite{A.E.Allahverdyan_2004}.\\

Let us now consider a bipartite quantum state $\rho^{AB}$. If the subsystem $B$ is discarded, the maximum amount of work extractable from the subsystem $A$ is simply equal to $\mathcal{E}(\rho^A)$, that is, the ergotropy of the reduced state $\rho^A = \hbox{Tr}_B[\rho^{AB}]$. Let us now assume that a measurement, described by POVM operators $\{\Pi_b\}$ is performed on the subsystem $B$ and that one can use the information obtained from this measurement to extract energy from $A$; in this scenario the maximum amount of extractable work via unitary operations is given by the daemonic ergotropy~\cite{francica},
\begin{align}
    \overline{\mathcal{E}}_{\{\Pi_b\}} = \sum_b p_b \,\mathcal{E}(\rho_b^A) \,,
    \label{eq:daemonicergo}
\end{align}
that is the average over the outcomes probability $p_b = {\rm Tr}[\rho^{AB} (\mathbb{I}\otimes \Pi_b)]$ of the ergotropies of the condtional states $\rho_b^A = {\rm Tr}_B[\rho^{AB} (\mathbb{I}\otimes \Pi_b)]$. Since ergotropy is a convex function,
the daemonic ergotropy is always larger than the ergotropy of the unconditional state: $    \overline{\mathcal{E}}_{\{\Pi_b\}}  \geq \mathcal{E}(\rho^A)$, mathematically quantifying the fact that a hypotetical Maxwell daemon can indeed extract more work by exploiting the information obtained from the measurement, and by adapting and optimizing the corresponding work extraction unitary for each conditional state. In~\cite{francica} it is shown how the enhancement in the work extraction ben be linked to the presence of quantum correlations, in particular to quantum discord, between the two parties. A more thorough study on the kind of correlations allowing to observe a daemonic gain is described in~\cite{Bernards2019}, along with an algorithm able to identify the optimal (not necessarily projective) POVM maximizing the daemonic ergotropy.\\

The concept of daemonic ergotropy has then been extended to the scenario of Markovian continuously monitored quantum systems~\cite{Morrone2023}. In this case, analogously to what was described in Sec.~\ref{s:gaussian_open} for Gaussian quantum states, one obtains a conditional dynamics, ruled by a stochastic master equations for the conditional state $\rho_c$~\cite{Albarelli2024}. Different measurements strategies yields different stochastic master equations, and thus different {\em unravellings} of the corresponding Markovian master equation for the unconditional quantum state $\rho_{\sf unc}=\mathbb{E}[\rho_c]$. The daemonic ergotropy for a specific unravelling is then simply defined as 
\begin{align}
\overline{\mathcal{E}}_{\sf unr}= \mathbb{E}[\mathcal{E}(\rho_c)]\,,
\end{align}
that is as the average ergotropy of the quantum trajectories $\rho_c$. In this framework, the convexity of the ergotropy allows to state that $\overline{\mathcal{E}}_{\sf unr}\geq \mathcal{E}(\rho_{\sf unc})$, that is the daemonic ergotropy is always larger or equal than the ergotropy of the unconditional state. As in the previous scenario, the interpretation behind this result is that a Maxwell daemon is able to optimize the work extraction for each quantum trajectory by exploiting the information obtained by measuring the environment. By studying some paradigmatic examples, it was also shown in~\cite{Morrone2023} how for mixed states unravellings a hierarchy between the different monitoring strategies exists, while in~\cite{Elyasi2025} an experimental simulation of a daemonic enhanced work extraction in the open quantum system scenario has been daemonstrated on a digital quantum computer. 
\section{Ergotropy of Gaussian states}
\label{s:ergo_gauss}
We intend to determine and study the ergotropy of Gaussian states with respect to quadratic Hamiltonians, which form the backbone of their description and are ubiquitous in physics. 

The most general derivation and formula for the ergotropy of Gaussian states given a generic quadratic Hamiltonian describing the energy of the system can be found in~\cite{HovhannisyanGaussErgo2020,rodriguez2025extractingenergybosonicgaussian}. However, because of the normal mode decomposition, any strictly positive Hamiltonian matrix can be turned, through a symplectic transformation (equivalent to a unitary operation at the Hilbert space level), into the ``free'' Hamiltonian $
\hat{H}_0$ in Eq.~\eqref{eq:freehamiltonian}.
Since, as we saw, any Gaussian state may be obtained by applying a symplectic transformation on a thermal state of the free Hamiltonian $\hat{\nu}= \otimes_j \hat{\nu}_j$ we can, 
without loss of generality, proceed and restrict ourselves to evaluate the ergotropy of any Gaussian state with respect to the free Hamiltonian set out above. Besides, this very 
same fact makes the evaluation of the ergotropy of Gaussian states immediate: as it is apparent from Eq.~\eqref{eq:rhoGauss}, the corresponding passive state is indeed $\tau_{\rho_G}=\hat{\nu}$
and the minimisation in Eq.~(\ref{eqn1: generalergo}) is in fact always achieved by the unitary corresponding to the symplectic transformation that turns the state's CM in Williamson normal form, followed by the displacement that sets its first moments to zero. 
Now, by first evaluating the energy of a thermal state $\hat{\nu}$ via Eq.~\ref{eq:energy} in terms of the symplectic eigenvalues $\nu_j$ as
\begin{align}
    E(\hat{\nu}) = \frac12 \sum_{j=1}^n \nu_j \,,
\end{align}
and then by exploiting the formula for the ergotropy in Eq.~\eqref{eq:ergo_passive}, one directly obtains the following closed formula for the ergotropy
\begin{equation}
\mathcal{E}(\rho_G) = \frac14 {\rm tr}\bm\sigma +\frac12 |\Bar{\bf r}|^2 - \frac12 \sum_{j=1}^n \nu_j .
\end{equation}
\subsection{Single-mode Gaussian states}

Henceforth, we shall focus on the relevant case of single-mode systems. For single-mode Gaussian states, $\nu_1=\sqrt{{\rm Det}\bm\sigma}$, which follows from the fact that ${\rm Det}S=1$ for all $S\in Sp_{2n}({\mathbb R})$.
Hence, the ergotropy of a single-mode Gaussian state is given by
\begin{align} \label{ergo1m}
\mathcal{E}(\rho_G) &= \frac12 |\Bar{\bf r}|^2 + \frac14 {\rm tr}\bm\sigma - \frac12 \sqrt{{\rm det}\bm\sigma}\,, \\
&= E(\rho_G) - \frac1{2\mu(\rho_G)} \,,
\end{align}
where in the second line we have shown the explicit dependence of the ergotropy on the energy and on the purity of the Gaussian state. One should notice that a similar, nonetheless less explicit, behaviour can be observed also for a qubit quantum state $\rho_{\sf qubit}$, whose ergotropy can be written as~\cite{morrone2023charging}
\begin{align}
\mathcal{E}(\rho_{\sf qubit}) = E(\rho_{\sf qubit}) + \frac1{2}\sqrt{2\mu(\rho_{\sf qubit})-1} \,.
\end{align}

Now, the first and second moments of a state are completely independent, and so are their contributions to the ergotropy.
Not surprisingly, the contribution of the first moments just grows monotonically with their magnitude.


It is also instructive to analyse the contribution to the ergotropy of each real parameter characterising the CM. Because of the SVD of symplectic transformations and of the fact that the normal form is always proportional to the identity (and thus rotationally invariant), the most general single-mode CM may be written as $\bm\sigma=\nu R_\varphi Z^2 R_{\varphi}^\top$, where $\nu\ge 1$ is the symplectic eigenvalue, $Z={\rm diag}(z,1/z)$ for $z>0$ and 
\begin{equation}\label{rota}
R_\varphi = \left(\begin{array}{cc}
\cos\varphi&\sin\varphi\\
-\sin\varphi&\cos\varphi\end{array}
\right),
\end{equation}
for $\varphi\in[0,2\pi[$. The ergotropy is then 
\begin{equation}
\mathcal{E} = \frac14 \nu(z^2+\frac{1}{z^2}-2) + \frac12 |\Bar{\bf r}|^2 .
\label{eq:ergotropyGauss0}
\end{equation}
The ergotropy is thus independent from the optical phase $\varphi$ and increasing with both squeezing $(z^2+1/z^2)$ and thermal noise $\nu$. Notice however that, as one should expect, for no squeezing ($z=1$) the ergotropy has zero contribution from the thermal noise $\nu$, since the term multiplying it vanishes. Remarkably we observe that, as long as any squeezing operation is present in the symplectic diagonalization of the covariance matrix, the thermal noise amplifies, so to speak, the energy extraction process. 
\section{Daemonic ergotropy of Gaussian states}
\label{s:dergo_gaussian_measured}
Let us now consider a generic multipartite Gaussian state, partitioned into subsystems A and B and described by first moment vector $\Bar{\bf r}$ and CM $\bm\sigma$ as in Eq.~\eqref{eq:2ms}. 
Our first goal is to derive the daemonic ergotropy corresponding to the scenario where a general-dyne measurement $\bm\sigma_{\sf m}$ is performed on party $B$ and work is extracted from party $A$, and thus one obtains conditional states described by first moment vector and CM reported in Eqs.~\eqref{eqn:gendymapR} and~\eqref{eqn:gendynemapCM}. We start by observing that the formula for the daemonic ergotropy in Eq.~\eqref{eq:daemonicergo} can be rewritten via Eq.~\eqref{eq:ergo_passive} as
\begin{align}
    \overline{\mathcal{E}}_{\{\Pi_b\}} &= \sum_b p_b\left( E(\rho_b^A)- E(\tau_b^A) \right)\,, \\
    &= E(\rho^A) - \sum_b p_b \,E(\tau_b^A) \,,
\end{align}
in terms of the passive states $\tau_b^A$  corresponding to the conditional states $\rho_b^A$. Since the CM of the conditional states is deterministic, and the passive state of a Gaussian state is completely determined by the CM, the second term can be readily evaluated and one can write a closed formula for the daemonic ergotropy for Gaussian states as 
\begin{align}
    \overline{\mathcal{E}}_{\bm\sigma_{\sf m}} = 
    \frac14 {\rm tr}\bm\sigma_A +\frac12 |\Bar{\bf r}_A|^2 - \frac12 \sum_{j=1}^{n_A} \nu_{A,j}^{(c)} \,,
    \label{eq:daemonicergoG0}
\end{align}
where we denote with $\nu_{A,j}^{(c)}$ the symplectic eigenvalues of the (deterministic) CM $\bm\sigma_A^{(c)}$ of the conditional states we reported in Eq.~\eqref{eqn:gendynemapCM}. We observe here that as expected for no correlations $\bm\sigma_{AB}=0$, that is for a product Gaussian state $\rho^{AB} = \rho^A \otimes \rho^B$, one obtains $\overline{\mathcal{E}}_{\bm\sigma_{\sf m}} = \mathcal{E}(\rho^A)$ and thus no daemonic gain can be observed. For Gaussian states, non-vanishing correlations $\bm\sigma_{AB}\neq0$ is a necessary and sufficient condition for non-zero discord~\cite{adessodatta,GaussDiscordGiordaParis} and non-zero quantum mutual information, and, if we restrict to pure states, for non-zero entanglement. We have thus proven that (quantum) correlations are a necessary condition for observing a daemonic gain $\overline{\mathcal{E}}_{\bm\sigma_{\sf m}} > \mathcal{E}(\rho^A)$. This result for Gaussian states strengthens the findings of~\cite{francica}, where it was proven that non-zero discord is sufficient for daemonic gain, but the converse does not hold in general. While this result seems to give a particular emphasis on the relationship between daemonic ergotropy and correlations, in Sec.~\ref{s:tmsts} through specific examples we will actually show that no quantitative (monotonous) relationship exists between these quantities, as in several instances ergotropy and (quantum and classical) correlations show a opposite behaviour as a function of certain physical parameters.

If we now consider the situation where subsystem $A$ corresponds to a single mode, then the formula can be directly written as
\begin{align}
    \overline{\mathcal{E}}_{\bm\sigma_{\sf m}} =& 
    \frac14 {\rm tr}\bm\sigma_A +\frac12 |\Bar{\bf r}_A|^2 - \frac12 \sqrt{{\rm det}[\bm\sigma_A^{(c)}]} \,, \\
    =&\frac14 {\rm tr}\bm\sigma_A +\frac12 |\Bar{\bf r}_A|^2 \nonumber\\
    &- \frac12 \sqrt{{\rm det}[\bm\sigma_A-\bm\sigma_{AB}\bm(\bm\sigma_B+\bm\sigma_{\sf m})^{-1}\bm\sigma_{AB}^{\top}]} \,.
    \label{eq:daemonicergoG}
\end{align}
We also notice that the formula above can be conveniently rewritten as
\begin{align}
    \overline{\mathcal{E}}_{\bm\sigma_{\sf m}} &= E(\rho^A) - \frac1{2 \mu[\rho_c^A]}\,,
    \label{eq:daemonicergoG_purity}
\end{align}
that is in terms of the energy $E(\rho^A)$ of the reduced Gaussian state of the subsystem A, and of the purity $\mu[\rho_c^A]$ of the single-mode conditional states (we remind that, as the conditional CM is deterministic, the purity of each conditional state is deterministic too). It is then clear that, for single-mode Gaussian systems, larger daemonic ergotropies are obtained via measurements on correlated subsystems that lead to better purification (cooling) of the conditional states.
\subsection{General-dyne optimization of the daemonic ergotropy}
We now restrict to the scenario of two-mode Gaussian states, and look for the maximisation of the function $\overline{\mathcal E}_{\bm\sigma_{\sf m}}$ in Eq.~\eqref{eq:daemonicergoG} over the physical CM $\bm\sigma_{\sf m}$ that characterises the general-dyne daemon acting on subsystem $B$.

Let us first recall that for pure initial bipartite state, i.e., such that ${\rm det}\bm\sigma_{AB}=1$, then any rank-one general-dyne measurements, that is such that ${\rm det}\bm\sigma_{m}=1$ is going to be optimal~\cite{Bernards2019,Morrone2023}, as all the conditional states are pure and one simply obtains that the optimized daemonic ergotropy is equal to the energy of the reduced state (a part from a constant, corresponding to the energy of the ground state), $\overline{\mathcal{E}}_{\sf max}= E(\rho^A) - 1/2= \frac14 {\rm tr}\bm\sigma_A +\frac12 |\Bar{\bf r}_A|^2$.

So, let us now focus on initial mixed bipartite states, where the measurement optimization is not trivial. First off, notice that the optimisation over all general-dyne measurements on mode $B$ allows us to apply any symplectic on it without loss of generality: we can therefore put $\bm\sigma_B$ in Williamson normal form, $\bm\sigma_B=b\openone_2$ and then apply a further proper rotation, which allows us to undo a rotation in the SVD of $\bm\sigma_{AB}$, so as to write $\bm\sigma_{AB}=R_{\eta}{\rm diag}(c_+,c_-)$, where $|c_+|$ and $|c_-|$ are the singular values of the completely generic matrix $\bm\sigma_{AB}$ (it should be noted that one can set, without loss of generality, $|c_+|\ge|c_-|$, $c_+\ge0$, and $c_-\ge 0$ if ${\rm Det}{\bm\sigma_{AB}}\ge 0$ or $c_-<0$ if ${\rm Det}{\bm\sigma_{AB}}<0$).

Furthermore, we can take advantage of the fact that the ergotropy of the conditional states of mode $A$ (after the measurement) is invariant under phase rotations, and diagonalise the matrix $\bm\sigma_A=a{\rm diag}(z,1/z)$ (the action of this rotation on $\bm\sigma_{AB}$ can be absorbed by redefining the angle $\eta$ introduced above).

Summing up we can, without loss of generality, consider an initial two-mode state of the form (\ref{eq:2ms}) 
with submatrices
\begin{equation}\label{eq:symmp}
\bm\sigma_A=a\,{\rm diag}(z_A,1/z_A) , \;
\bm\sigma_B=b\openone_2 , \;
\bm\sigma_{AB}=R_{\eta}{\rm diag}(c_+,c_-) .
\end{equation}
We also remark that, since $1$ vs $n$ mode Gaussian states may always be turned into $1$ vs $1$ mode states through a symplectic operation \cite{SERAFINI_2023}, our argument applies to the daemonic ergotropy of $1$ vs $n$ mode states too.

We have thus reduced our analysis to a problem with $6$ real parameters (as well as the initial first moments, whose roles are however trivial, as we shall see). While the initial squeezing $z_A$ and the angle $\eta$ are unconstrained, $a$, $b$, $c_+$ and $c_-$ are bound by the Heisenberg uncertainty relation as follows:
\begin{equation}
(ab-c_+^2)(ab-c_-^2)-a^2-b^2-2c_+c_-+1 \ge 0,
\end{equation}
as well as by $a\ge1$, $b\ge1$ and $|c_+|\ge |c_-|$ (the latter stemming from the free ordering allowed by the SVD).

The CM describing the general-dyne measurement on the single-mode subsystem $B$ can be generally written as 
$\bm\sigma_{\sf m}=\nu_m R_{\vartheta_{\sf m}}{\rm diag}(z_{\sf m},1/z_{\sf m})R_{\vartheta_{\sf m}}^{\top}$. 
As already discussed, the ergotropy is maximised by maximising the conditional purity ${\rm Tr}[
(\rho_c^{A})^2]$. Eq.~(\ref{eqn:cmpovm}) shows that any non-ideal ($\nu_m>1$) general-dyne measurement results into a statistical mixture of Gaussian states with the same second moments as those resulting from an ideal ($\nu_m=1$) general-dyne (note that, as well known, the conditional second moments are deterministic for Gaussian measurements, i.e., they do not depend on the measurement outcome) and different first moments (because of the integral over ${\rm d}{\bf y}$). Hence, the convexity of the purity guarantees that $\nu_m=1$ achieves the optimal puritification and is thus optimal to our aims. 

Let us now focus on the other parameters characterizing the measurement: $\vartheta_{\sf m}$ is an angle between zero and $\pi/2$, while in principle $z_{\sf m}$ takes values from zero to infinity; however, it is easy to observe that one can map $z_{\sf m}$ to $1/z_{\sf m}$ by a subsequent rotation, and thus by properly changing the value of $\vartheta_{\sf m}$. For this reason we can restrict the parameters range to $z_{\sf m} \in ]0,1]$ and $\vartheta_{\sf m} \in [0,\pi/2]$. 

The optimisation over the general-dyne angle may be obtained analytically by minimising the determinant entering Eq.~(\ref{eq:daemonicergoG}), yielding the optimal general-dyne phase 
\begin{widetext}
\begin{equation}
    \vartheta_{\sf m,opt} = \frac{\pi}{2}+\frac12\arctan\left(\frac{2\sin(2\eta)c_+c_-(z_A^2-1)}{(c_+^2-c_-^2)(z_A^2+1) - \cos(2\eta)(c_+^2+c_-^2)(z_A^2-1)}\right).
\end{equation}
This value of $\vartheta_{\sf m,opt}$ then yields the daemonic ergotropy for two-mode Gaussian states maximised 
over general-dyne measurements:
\begin{align}\label{genmax}
    \overline{\mathcal{E}}_{\sf Gen,max}= 
    &\frac12|\Bar{\bf r}_A|^2 + 
    \frac14(z_A+\frac{1}{z_A})a 
    -  \frac12\bigg(4c_+^2c_-^2z_Az_{\sf m} -(c_+^2+c_-^2)(1+z_A^2)(1+2bz_{\sf m}+z_{\sf m}^2)a+ 4 z_A (b + z_{\sf m}) (1 + b z_{\sf m}) a^2 \nonumber\\
    &+ (z_A^2-1) a \cos(2\eta) \left((c_+^2 - c_-^2)  (1 + 2 b z_{\sf m} + z_{\sf m}^2) - (c_+^2 + c_-^2) (z_{\sf m}^2-1) \cos(2 \vartheta_{\sf m,opt})\right) \nonumber \\ &+(z_{\sf m}^2-1) a \big((c_+^2 - c_-^2) (z_A^2+1) \cos(2 \vartheta_{\sf m,opt}) + 
      2 c_+ c_- (z_A^2-1) \sin(2 \eta) \sin(2 \vartheta_{\sf m,opt})\big)\bigg)^{\frac12}/\sqrt{4 z_A (b + z_{\sf m}) (b z_{\sf m} + 1)}.
\end{align}

It is worthwhile to report the homodyne limit $z_{\sf m}\rightarrow 0$:
\begin{align}
    \overline{\mathcal{E}}_{\sf Hom,max}=& 
    \frac12|\Bar{\bf r}_A|^2 +
    \frac14(z_A+\frac{1}{z_A})a 
    -  \frac12\bigg(a^2-\frac{a\sin^2\eta}{bz_A}(z_A^2c_+^2\cos^2\vartheta_{\sf m,opt}+c_-^2\sin^2\vartheta_{\sf m,opt})
    \nonumber\\
    &-\frac{a\cos^2\eta}{bz_A}(c_+^2\cos^2\vartheta_{\sf m,opt}+z_A^2c_-^2\sin^2\vartheta_{\sf h,opt})
    -\frac{a\sin(2\eta)}{2bz_A}(z_A^2-1)c_+c_-\sin(2\vartheta_{\sf m,opt})\bigg)^{\frac12},
\end{align}
as well as the heterodyne case $z_{\sf m}=1$ (where there is no dependence on $\vartheta_{\sf m}$):
\begin{align}\label{Ehet}
    \overline{\mathcal{E}}_{\sf Het} =& 
    \frac12|\Bar{\bf r}_A|^2 +
    \frac14(z_A+\frac{1}{z_A})a 
     - \frac{1}{\sqrt{z_A}(1 + b)} 
\left[ 
  c_{-}^2 c_{+}^2 z_A \cos^4(\eta)  \right. 
  \left. \,+\,
  \left( a(1 + b) z_A - c_{-}^2 \sin^2(\eta) \right) 
  \left( a + a b - c_{+}^2 z_A \sin^2(\eta) \right) \right.  \\
  & \left. \,+\, 
  \cos^2(\eta) \left( 
    -a(1 + b)(c_{+}^2 + c_{-}^2 z_A^2) \nonumber \right. \right. 
     \left.\left. +\, 2 c_{-}^2 c_{+}^2 z_A \sin^2(\eta) 
  \right) 
\right]^{1/2} .
\end{align}
\end{widetext}

There exist cases whose phase-dependence is such that the optimal measurement is a general-dyne measurement with finite $z_{\sf m}\neq1$. It is easy to optimise the expression (\ref{genmax}) with respect to $z_{\sf m}$ in each specific instance. The most prototypical class of two-mode entangled Gaussian states can be dealt with exactly, and deserves a dedicated treatment.

\subsection{Example: maximal daemonic ergotropy of phase-invariant and two-mode squeezed thermal states}
\label{s:tmsts}

The case of phase-invariant states, which in our parametrisation corresponds to $c\equiv c_1=\mp c_2$, $z_A=1$ and $\eta=0$, is particularly relevant, as it subsumes thermally seeded two-mode squeezed states, which describe non-degenerate parametric down conversion at finite, non-zero temperature. 

In all such cases, the conditional determinant ${\rm det}\bm\sigma_A^{(c)}$, which we are tasked to minimise so as to maximise the conditional purity and hence the ergotropy, reads simply $${\rm Det}\bm\sigma_{A,PI}^{(c)}= \frac{(a + a b z_{\sf m} - c^2 z_{\sf m}) (-c^2 + a (b + z_{\sf m}))}{(b + z_{\sf m}) (1 + b z_{\sf m})}.$$ The derivative of this quantity with respect to $z_{\sf m}$ is proportional to a positive factor by $(z_{\sf m}^2-1)$. Therefore, the heterodyne daemon, and the associated expression $\overline{\mathcal{E}}_{\sf Het}$ of Eq.~(\ref{Ehet}), is always optimal in phase-invariant cases. We remark that (efficient) heterodyne, and in fact any (efficient) general-dyne detection with $z_m\neq \{ 0,\infty \}$ (not homodyne) does not correspond to a projective measurement, but to a rank-one generalized (POVM) measurement (over a set of continuously-parametrised, non-orthogonal states). We have thus found that also for continuous-variable system generalized measurements may allow to obtain higher values of daemonic ergotropy, as already demonstrated in discrete-variable systems~\cite{Bernards2019}.

For thermally seeded two-mode squeezed states, with squeezing parameter $r$ acting on a thermal state with a mean number $N$ of excitations, one has set of values $a=b=(2N+1)\cosh(2r)$, $c_+=-c_-=(2N+1)\sinh(2r)$, $z_A=1$ and $\eta=0$. Then, the maximal (heterodyne) daemonic ergotropy reads
\begin{align}
    \overline{\mathcal E}^{\sf (2ms)}_{\sf Het} = \frac{(2N+1)^2 \sinh^2 (2r)}{2 + 2 (2N+1) \cosh(2r)} .
\end{align}
As regards homodyne detection, one observe that, due to phase-invariance, there is not a privileged value for the homodyne phase, and one gets the result
\begin{equation}
\overline{\mathcal E}^{\sf (2ms)}_{\sf Hom,max} = (2N+1) \sinh^2 r .
\end{equation}
Of course, $\overline{\mathcal E}^{\sf (2ms)}_{\sf Het,max} \geq \overline{\mathcal E}^{\sf (2ms)}_{\sf Hom,max}$, with the two ergotropies being equal as expected only for $N=0$: indeed, for globally pure states, any pure general-dyne detection, corresponding to the projection on rank-1 projectors, will result into a pure local conditional state.

We also observe that, as the reduced state $\rho^A$ is a thermal state, the ergotropy if no measurement is performed on subsystem $B$ would be equal to zero. On the other hand both daemonic ergotropies are monotonically increasing with the number of thermal excitations $N$.
This shows that general-dyne daemons may turn heat into energy, as long as any degree of correlations is present; in fact, as we observed in Eq.~\eqref{eq:ergotropyGauss0}, the presence of a squeezing operation in the symplectic diagonalization of the conditional states, the thermal noise amplifies the extracted work. After Eq.~\eqref{eq:daemonicergoG0} we have observed how non-zero daemonic gain is equivalent to non-zero entanglement for pure states, and to non-zero discord and non zero quantum mutual information for mixed states. We can now investigate in this example whether a quantitative relationship exists between daemonic ergotropy and quantum correlations. For such states we can analytically quantify
entanglement (logarithmic negativity, as per \cite{vidal02,SERAFINI_2023}), quantum mutual information and quantum discord \cite{adessodatta,GaussDiscordGiordaParis,SERAFINI_2023}. One may then notice that as expected daemonic work may be extracted through marginal measurements even if the original state is not entangled, as is the case for ${\rm e}^{2|r|}\le (2N+1)$. Also, although all thermally-seeded two-mode squeezed states have non-zero discord and non-zero quantum mutual information, both these quantities are monotonically decreasing with the thermal noise $N$ (as well as entanglement), and thus exhibit a behaviour opposite to that of the daemonic ergotropy, disproving any monotonous relationship between daemonic ergotropy and (quantum and classical) correlations.
\section{Daemonic ergotropy for continuously monitored Gaussian systems}
\label{s:dergo_gaussian_open}
Let us now address the problem of evaluating the daemonic ergotropy 
for open Markovian Gaussian systems continuously monitored via general-dyne measurements, 
that we described in Sec.~\ref{s:gaussian_open}.
The scenario is fundamentally similar to the one we have just described, as
also in this case we will observe a probabilistic (stochastic) evolution 
of the first moments, and a deterministic evolution of the CM of the 
conditional states. As a consequence we can generally evaluate the 
daemonic ergotropy, given a certain open system dynamics (described by
the coupling matrix $C$ and the bath CM $\bm\sigma_{\sf in}$) and a
general-dyne measurement CM $\bm\sigma_{\sf m}$ via the formula
\begin{align}
    \overline{\mathcal{E}}_{\bm\sigma_{\sf m}} = 
    \frac14 {\rm tr}\bm\sigma_{\sf unc} +\frac12 |\Bar{\bf r}_{\sf unc}|^2 - \frac12 \sum_{j=1}^{n} \nu_{j}^{(c)} \,,
\end{align}
where here we denote with $\nu_{j}^{(c)}$ the symplectic eigenvalues of the (deterministic) CM $\bm\sigma_{c}$ of the conditional states evolving according to the Riccati equation~\eqref{eq:riccati}, and where we remind that $\bm\sigma_{\sf unc}$ and $\Bar{\bf r}_{\sf unc}$ correspond respectively to the CM and the first moment vector of the unconditional state evolving according to Eqs.~\eqref{eqn1: unconditionalfirst} and ~\eqref{eqn1 unconditional dynamics second}. For a single-mode system, then the formula can be directly written as
\begin{align}
    \overline{\mathcal{E}}_{\bm\sigma_{\sf m}} &= 
    \frac14 {\rm tr}\bm\sigma_{\sf unc} +\frac12 |\Bar{\bf r}_{\sf unc}|^2 - \frac12 \sqrt{{\rm det}\bm\sigma_c} \,, \\
    &=E(\rho_{\sf unc}) - \frac1{2\sqrt{\mu[\rho_{c}]}} \,,
    \label{eq:daemonicergo_monitoring}
\end{align}
where again it becomes clear how larger daemonic work extraction does
actually correspond to unravellings that are able to better purify 
the corresponding conditional states. Although this situation is analogous to the maximisation performed in the previous section, in this case we have a proper dynamics  which actually depends on the measurement we have chosen to perform. 
As before we will now proceed by discussing how to identify the optimal general-dyne, maximising 
the daemonic ergotropy for single-mode Gaussian systems; 
we will first consider and fully characterise the 
steady-state scenario, and we will then assess what happens in the transient regime
by focusing on a paradigmatic example.

We start by observing that, to obtain a steady-state value for $\overline{\mathcal{E}}_{\bm\sigma_{\sf m}}^{ss}$ one needs to have a steady-state solution for both the unconditional dynamics, $\bm\sigma_{\sf unc}^{ss}$ and $\Bar{\bf r}_{\sf unc}^{ss}$, and for the conditional CM $\bm\sigma_c^{ss}$. One can easily show that having a Hurwitz drift matrix $A$ (i.e. such that all the eigenvalues $\{\lambda\}$ of $A$ satisfy ${\rm Re}(\lambda) < 0$ ) is a necessary and sufficient condition to satisfy the requirements above: in fact $A$ being Hurwitz is a necessary and sufficient condition for the existence of $\bm\sigma_{\sf unc}^{ss}$ and $\Bar{\bf r}_{\sf unc}^{ss}$, and a sufficient condition for the existence of $\bm\sigma_c^{ss}$ (if $A$ is Hurwitz, then one has $(B,A)$-detectable by definition, see ~\cite{WisemanDoherty,BAO2024111622} for more details). Henceforth we will assume that this condition is satisfied for the dynamics we will consider. 

Before proceeding with the optimization of the general-dyne measurement, we first note that, without loss of generality, we can apply a symplectic operation on the environment (offset by considering any possible general-dyne detection in the optimisation) and take an environmental CM $\bm\sigma_{\sf in}=\nu_{\sf in}\openone_2$, for $\nu_{\sf in}\ge 1$ quantifying the amount of thermal noise in the environment. As mentioned above, the optimisation of the deamonic ergotropy over the general-dyne measurement $\bm\sigma_{\sf m}$ amounts to determine the optimal increase of purity of the conditional states, and thus to minimising the symplectic eigenvalue of $\bm\sigma_c^{ss}$. 
The argument that we applied in the previous section, regarding the convexity of the ergotropy in terms of the conditional CM $\bm\sigma_c$ still applies, and thus we can already state that the optimal unravelling will always correspond to an efficient general-dyne measurement. In the following we will show how the optimal general-dyne is not necessarily unique. We will now focus on the optimization at steady-state, differentiating between the case where the environment is at zero-temperature or at finite temperature.

{\em Steady-state optimization for a zero temperature environment -- }In this case we can set $\nu_{\sf in}=1$, and we are thus describing an environment in a pure state (e.g. describing pure loss dynamics, or the interaction with a pure squeezed bath). If we assume now that the system is inizialized in a pure state and that the continuous monitoring is a performed via an efficient general-dyne detection, that is such that ${\rm det}\bm\sigma_{\sf m}=1$ and corresponding in the Hilbert space to rank-one POVM operators, then the conditional state remains pure during the whole dynamics, and thus one also obtains ${\rm det}\bm\sigma_c^{ss}=1$. Since the conditional steady-state, if exists, is unique and obtained irregardless of the initial state, we have just proved that any efficient general-dyne detection is going to completely purify all conditional states in the long time dynamics. Thus one obtains that for zero temperature environments the optimized steady-state daemonic ergotropies is simply equal to the unconditional steady-state energy (a part from the zero point energy) $\overline{\mathcal{E}}_{\sf max}^{ss}=E(\rho_{\sf unc}^{ss})-1/2$ and is achieved for any general-dyne measurement such that $\det\bm\sigma_{\sf m}=1$.  

{\em Steady-state optimization for a non-zero temperature environment -- }
In the presence of a thermal environment, we can set the noise parameter $\nu_{\sf in}>1$; let us now focus on efficient homodyne detection, such that we can write the matrix
$(\bm\sigma_{\sf in} +\bm\sigma_{\sf m})^{-1}$, key to the parametrisation of our monitoring, as 
\begin{align}
(\bm\sigma_{\sf in} +\bm\sigma_{\sf m})^{-1} =& \lim_{z_{\sf m}\rightarrow 0} R_{\vartheta_{\sf m}}{\rm diag}(\nu_{\sf in}+z_{\sf m} ,\nu_{\sf in}+1/z_{\sf m})^{-1} R_{\vartheta_{\sf m}}^{\top} \nonumber \\
=&  R_{\vartheta_{\sf m}} {\rm diag}(1/\nu_{\sf in} ,0) R_{\vartheta_{\sf m}}^{\top} ,
\end{align}
where $\vartheta_{\sf m}$ is the homodyne detection optical phase.
We recall that by setting the time-derivative in (\ref{eq:riccati}) to zero, we can write the Riccati equation for the steady-state conditional CM as
\begin{equation} \label{eqn1: algebraicricatti}
    \Tilde{A} \bm\sigma_c^{ss} + \bm\sigma_c^{ss}\Tilde{A}^\top+\Tilde{D}-\bm\sigma_c^{ss} BB^\top \bm\sigma_c^{ss} = 0 ,
\end{equation}
where $\Tilde{A} = A + E B^\top$ and $\Tilde{D}=D-E E^\top$. Since, for efficient homodyne measurements, the matrix $(\bm\sigma_{\sf in} +\bm\sigma_{\sf m})^{-1}$ scales as $1/\nu_{\sf in}$, one can directly estimate the scaling of the Riccati equation coefficients with $\nu_{\sf in}$ for pure general-dyne monitoring, and then evaluate the determinant of the general solution. One can easily verify from their definitions that $\tilde{A}$ does not depend on $\nu_{\sf in}$, while $\tilde{D}$ scales like $\nu_{\sf in}$ and $BB^\top$ scales like $1/\nu_{\sf in}$. Therefore, the solution to the Riccati equation must be $\bm\sigma_c^{ss}=\nu_{\sf in} \bm\sigma'$, where $\bm\sigma'$ represents the solution obtained for $\nu_{\sf in}=1$, which, as we have proven above, has determinant equal to $1$. This thus corresponds to obtain ${\rm det}[\bm\sigma_c^{ss}]=\nu_{\sf in}^2 = {\rm det}[\bm \sigma_{in}]$ and thus to the following formula for the optimal daemonic ergotropy under efficient homodyne monitoring at steady-state
\begin{equation}
\overline{\mathcal E}_{\sf Hom,max}^{ss}= \frac12|\Bar{\bf r}_{\sf unc}^{ss}|^2 +\frac14{\rm tr}\bm\sigma_{\sf unc}^{ss} -\frac{1}{2}\sqrt{{\rm det}\bm \sigma_{in}} \,,
\end{equation}
where $\Bar{\bf r}_{\sf unc}^{ss}=-A^{-1}{\bf d}$ and $\bm\sigma_{\sf unc}^{ss}$ solves the linear Lyapunov equation (\ref{eqn1 unconditional dynamics second}) with time-derivative set to zero. Notice that this result is completely general, applying to all couplings $C$ and local Hamiltonians $H_S$. 
As it turns out, at steady-state the optimal homodyne monitoring does not depend on the
homodyne angle, even in systems which are not phase-invariant: all efficient homodyne measurements will lead to conditional states having the same purity of the interacting environment.

However, the argument regarding the scaling of the matrix $(\bm\sigma_{\sf in} +\bm\sigma_{\sf m})^{-1}$ in terms of $\nu_{\sf in}$ cannot be extended to all efficient general-dyne detections. For this reason one will have to evaluate the steady-state conditional CM under general-dyne detection, and then identify case by case which is the optimal general-dyne unravelling.
\subsection{Example: daemonic ergotropy in a continuously monitored optical parametric oscillator}
It is interesting to assess the advantage allowed by monitoring in the interesting case of a optical parametric oscillator (producing single-mode squeezing) under loss and thermal noise, whose unconditional dynamics is described by the following Markovian master equation
\begin{align}
\frac{d\rho}{dt} &= -i \frac{\chi}{2}\left[\hat{x}\hat{p}+\hat{p}\hat{x},\rho \right] + \kappa (n_{th}+1)\mathcal{D}[\hat{a}]\rho \nonumber \\
&\,\,\, + \kappa \, n_{th}\mathcal{D}[\hat{a}^\dag]\rho \,,
\end{align}
where we have defined $\hat{a} = (\hat{x}-i\hat{p})/\sqrt{2}$ and the Lindblad superoperator $\mathcal{D}[\hat{A}]\rho = \hat{A}\rho\hat{A}^\dag - (\hat{A}^\dag \hat{A} \rho + \rho \hat{A}^\dag \hat{A})/2$. The physical parameters ruling the dynamics are $\chi$, describing the strength of the squeezing Hamiltonian, the loss parameter $\kappa$ and the number of bath thermal excitations $n_{th}$. The unconditional Gaussian dynamics is described by the drift matrix $A={\rm diag}
(-\kappa/2 - \chi, -\kappa/2 + \chi)$ and $D=\kappa \bm\sigma_{\sf in}=\kappa \nu_{\sf in} \openone_2$, where we have introduced $\nu_{\sf in}=(2 n_{th}+1)$~\cite{Diffusone,Albarelli2024}. 
Notice that stability (i.e. the drift matrix $A$ being Hurwitz) dictates $|\chi|<\kappa/2$ for this dynamics. For the sake of simplicity we will restrict ourselves to the range of positive  couplings, i.e. $0\leq \chi <\kappa/2$. The Gaussian steady-state can be easily derived and is characterized by a null first moment vector $\Bar{\bf r}_{\sf unc}^{ss}=0$ and a diagonal covariance matrix 
\begin{align}
\bm\sigma_{\sf unc}^{ss}= 
\nu_{\sf in} \begin{pmatrix}
    \frac{1}{1 + \tilde\chi} & 0 \\
    0 & \frac{1}{1-\tilde\chi} 
\end{pmatrix}\,
\label{eq:OPA_unconditionalCM}
\end{align}
where we have introduced the dimensionless parameter $\tilde\chi=2 \chi/\kappa$. The formula above shows how a squeezed $\hat{x}$ quadrature can be  obtained at steady-state, $(\bm\sigma_{\sf unc}^{ss})_{11}=2 (\Delta\hat{x}^2)<1$, for example by considering a zero-temperature environment $\nu_{\sf in}=1$ and for any value $0<\chi<\kappa/2$. 
The corresponding uncondtional steady-state ergotropy reads
\begin{equation}
{\mathcal E}_{\sf unc}^{ss} = \frac{\nu_{\sf in}}{2}\left(\frac{1}{1-\tilde\chi^2}-\frac{1}{\sqrt{1-\tilde\chi^2}}\right) .
\end{equation}
We can now derive the conditional states corresponding to the different unravellings; we will start by focusing only on homodyne and heterodyne strategies. 
In particular as regards to homodyne, by fixing the homodyne phase to $\vartheta_{\sf m} =0$ we obtain
\begin{align}
\bm\sigma_{\sf c,(Hom,\vartheta_{\sf m}=0)}^{ss}= 
\nu_{\sf in} \begin{pmatrix}
    1-\tilde\chi & 0 \\
    0 & (1-\tilde\chi)^{-1}
\end{pmatrix}\,,
\end{align}
leading to the minimum fluctuations in the $\hat{x}$ quadrature, and thus to maximum amount of squeezing in the $\hat{x}$ quadrature (infinite squeezing near instability, that is $\langle\Delta \hat{x}^2\rangle \to 0$ for $\tilde\chi\to 1$). 
On the other hand for homodyne phase $\vartheta=\pi/2$, one obtains 
\begin{align}
\bm\sigma_{\sf c,(Hom,\vartheta_{\sf m}=\pi/2)}^{ss}= 
\nu_{\sf in} \begin{pmatrix}
    (1+\tilde\chi)^{-1} & 0 \\
    0 & 1+\tilde\chi
\end{pmatrix}\,,
\end{align}
which may still yield squeezing in the $\hat{x}$ quadrature for $\nu_{\sf in}<(1+\tilde\chi)$, but much lower than what was obtained for $\vartheta_{\sf m}=0$. 
If we now consider heterodyne detection, we find
\begin{align}
\bm\sigma_{\sf c,Het}^{ss}& = {\rm diag}(\sigma^{\sf Het}_{11},\sigma^{\sf Het}_{22}) \nonumber \\
\sigma^{\sf Het}_{11}&= \frac{1}{2} \Big( 
  \nu_{\sf in} -1 - \tilde{\chi}(1 + \nu_{\sf in}) \nonumber \\
&   + 
  \sqrt{ 
    (1 + \nu_{\sf in}) \left[ 
      \nu_{\sf in} ( -1 + \tilde{\chi} )^2 + (1 + \tilde{\chi})^2 
    \right]  
  } 
\Big) \nonumber \\
\sigma^{\sf Het}_{22}&= \frac{1}{2} \Big( 
\nu_{\sf in}-1 + \tilde{\chi}(1 + \nu_{\sf in}) \nonumber \\
&  + 
  \sqrt{ 
    (1 + \nu_{\sf in}) \left[ 
      (-1 + \tilde{\chi})^2 + \nu_{\sf in} (1 + \tilde{\chi})^2 
    \right] 
  } 
\Big)\,
\label{eq:OPA_heterodyneCM}
\end{align}
which for a zero-temperature environment simplifies to $\bm\sigma_{\sf c,Het}^{ss} = {\rm diag}(-\tilde\chi + \sqrt{1+\tilde\chi^2},\tilde\chi + \sqrt{1+\tilde\chi^2})$. It is easy to check that in general heterodyne monitoring yields an intermediate value of squeezing in $\hat{x}$, respect to the two homodyne strategies. In the zero-temperature scenario, as we discussed before, despite the corresponding (pure) conditional states show very different features, the different unravellings lead to the same steady-state daemonic ergotropy, in formula
\begin{align}
\overline{\mathcal E}_{\sf max}^{ss} &= \frac{\tilde\chi^2}{2(1-\tilde\chi^2)} \,\,\,\,\,\,\,(\nu_{\sf in}=1)\,, \nonumber \\
&= \frac{2 \chi^2}{\kappa^2 - 4 \chi^2} \,\,\,\,\,\,\,(n_{\sf th}=0)\,.\label{eq:daemonicOPA_zeroTemp} 
\end{align}
For the non-zero temperature scenario ($\nu_{\sf in}>1$), as described above all homodyne unravellings still lead to the same steady-state unravelling, rescaled by the thermal noise, in formula
\begin{align}
\overline{\mathcal E}_{\sf Hom}^{ss} &=  \frac{\nu_{\sf in} \tilde\chi^2}{2(1-\tilde\chi^2)} \,, \nonumber \\
&= \frac{2 (2 n_{th} + 1) \chi^2}{\kappa^2 - 4 \chi^2} \,.
\label{eq:daemonicOPA_Homo}
\end{align}
In the formulas above we observe how the daemonic ergotropy is monotonically increasing with the thermal noise $n_{th}$, yielding another example where squeezing in the symplectic diagonalization of the conditional states, allow to exploit thermal noise for the work extraction.
Also for heterodyne detection, the corresponding daemonic ergotropy $\overline{\mathcal E}_{\sf Het}^{ss}$ can be evaluated analytically, by exploiting the formula in Eq.~\eqref{eq:daemonicergoG}, the unconditional covariance matrix in Eq.~\eqref{eq:OPA_unconditionalCM} and the heterodyne covariance matrix in Eq.~\eqref{eq:OPA_heterodyneCM}; as the formula is rather cumbersome, we are not reporting it here, but it is easy to check that
\begin{equation}
\overline{\mathcal E}_{\sf Het}^{ss} > \overline{\mathcal E}_{\sf Hom}^{ss}\,\,\, {\rm for} \,\, \nu_{\sf in}>1 \,,
\end{equation}
that is heterodyne is allowing to extract more work than any other homodyne unravelling whenever the bath has non zero temperature ($n_{th}>0$).

As regards the optimization over all the generaldyne measurement, one proves that if one fixes the measurement phase to zero, $\vartheta_{\sf m}=0$, then the conditional steady-state purity, and as a consequence the daemonic ergotropy is maximized for
the general-dyne parameter
\begin{align}
    z_{\sf m,opt} = \frac{1-\tilde\chi}{1+\tilde\chi} \,.
\end{align}
We have also numerically verified that non-zero phases $\vartheta_{\sf m}$ lead to worse result, and we can thus conjecture that this fully characterizes the optimal steady-state general-dyne monitoring. 
This result seems to suggest that, as one approaches the critical value $\tilde{\chi} \to 1$, the optimal general-dyne measurement tends toward homodyne detection, i.e., $z_{\sf m,opt} \to 0$. At first glance, this appears to contradict our previous findings, where we stated that heterodyne detection consistently outperforms homodyne.
\begin{figure}[tbp]
\centering 
\includegraphics[width=0.5\textwidth]{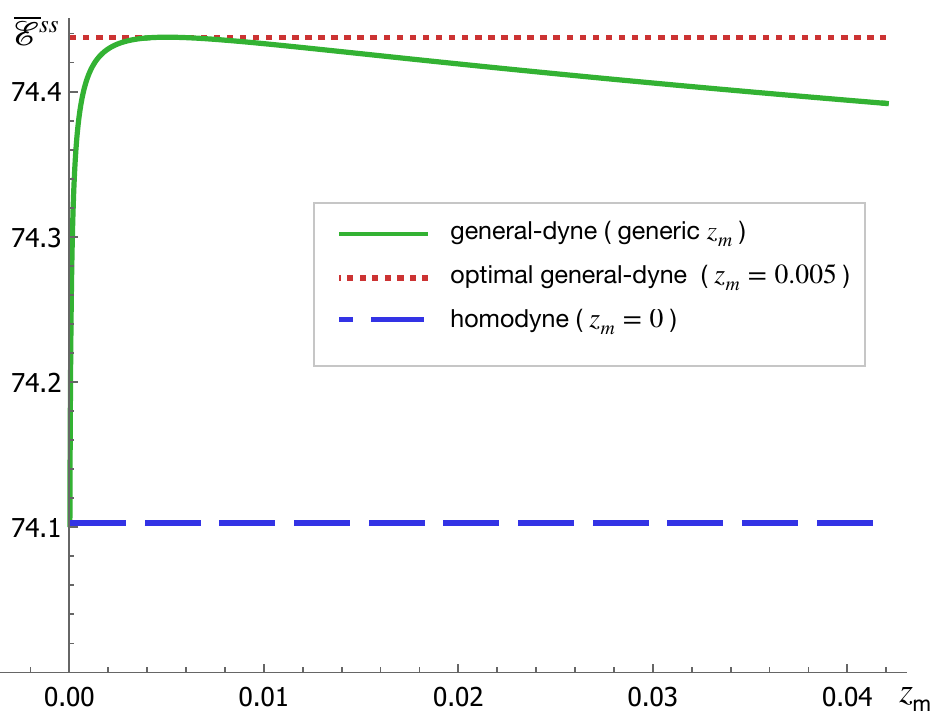}
\caption{(Color online) Steady-state daemonic ergotropy for a continuously monitored optical parametric oscillator interacting with a finite temperature environment ($\nu_{\sf in}=3$) and near criticality ($\tilde{\chi}=0.99$), as a function of the general-dyne parameter $z_{\sf m}$ (green solid line). The red dotted line corresponds to the maximum value obtained by fixing $z_{\sf m,opt}=0.005$, while the blue dashed line corresponds to the daemonic erogtropy obtained via general-dyne detection ($z_{\sf m}=1$).}
\label{f:daemonicOPA_GeneralDyne}
\end{figure}
However, as shown in Fig.~\ref{f:daemonicOPA_GeneralDyne}, for $\tilde{\chi} = 0.99$, there is a pronounced drop in $\overline{\mathcal{E}}^{ss}$ when varying $z_{\sf m}$ from its optimal value $z_{\sf m,opt}\approx 0.005$ to $z_{\sf m} = 0$, also confirming that $\overline{\mathcal{E}}_{\sf Het}^{ss} > \overline{\mathcal{E}}_{\sf Hom}^{ss}$. This behaviour is in fact confirmed
for any value of $\tilde{\chi}$.
As introduced previously, single-mode general-dyne detection can be implemented by generalizing the usual double-homodyne scheme needed for heterodyne detection: this is accomplished by choosing a beam-splitter transmissivity to achieve the desired value of $z_m$~\cite{GeneralDino}. However, the level of continuous fine-tuning necessary to implement the precise optimal general-dyne in experimental scenarios may go beyond current capabilities, in particular for beavhiours as the ones described here above. Crucially, enhancement is still obtainable: larger values of extractable work extraction is typically achievable compared to both the homodyne and heterodyne limits by choosing a sub-optimal detection setting within a wide region of $z_m$ values.\\

\begin{figure}[tbp]
\centering 
\includegraphics[width=0.5\textwidth]{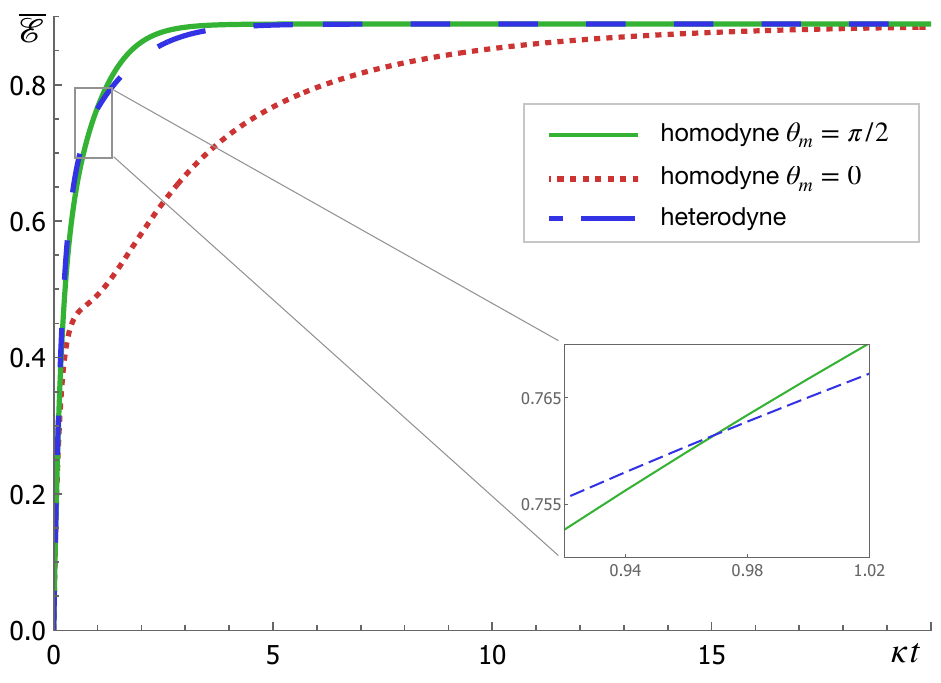}
\caption{(Color online) Daemonic ergotropy for a continuously monitored optical parametric oscillator interacting with a zero temperature environment, and continuously monitored respectively via homodyne detection with $\vartheta_{\sf m}=0$ (red dotted line), homodyne detection with $\vartheta_{\sf m}=\pi/2$ (green solid line) and heterodyne detection (blue dashed line). Parameters are fixed as follows: $\chi/\kappa=0.4$, $n_{th}=0$, $n_0=2$.
}

\label{f:daemonicOPA_ZeroTemp}
\end{figure}

We  now investigate what happens during the transient dynamics by considering as initial state a thermal state with $n_0$ thermal excitation, that is described by a CM $\bm\sigma_0 = (2 n_0 + 1) \openone_2$ and zero first moments. In this instance, we will focus on homodyne (with phases $\vartheta_{\sf m}=0$ and $\vartheta_{\sf m}=\pi/2$) and heterodyne only, being the more experimentally relevant scenarios. 
The results are shown in Fig.~\ref{f:daemonicOPA_ZeroTemp} for a zero temperature bath ($n_{th}=0$) and for $n_0=2$.
\begin{figure}[tbp]
\centering 
\includegraphics[width=0.5\textwidth]{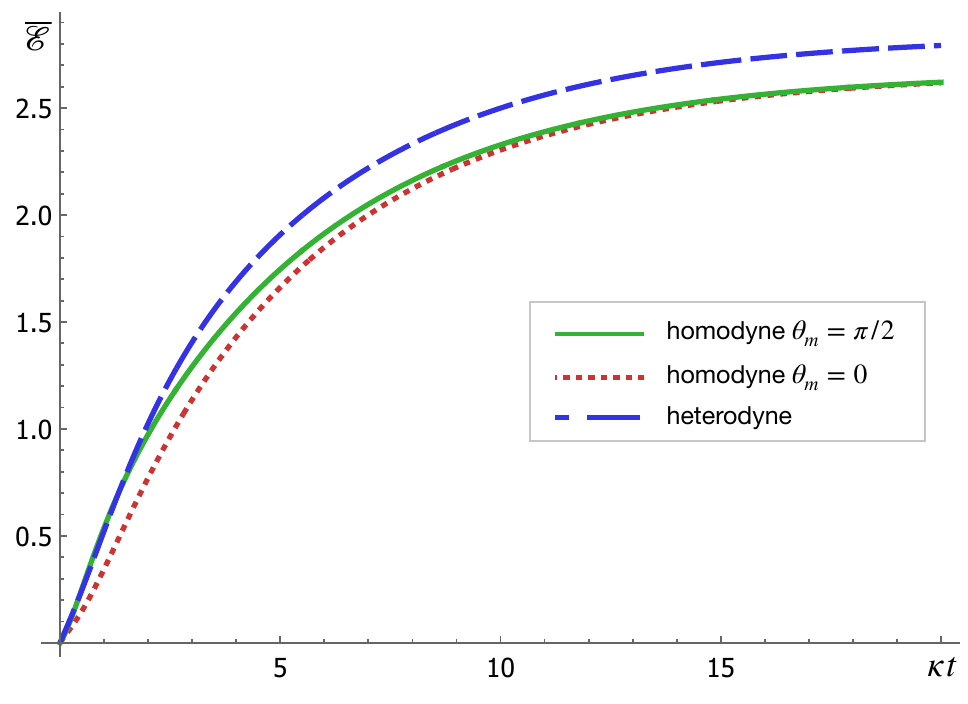}
\caption{(Color online) Daemonic ergotropy for a continuously monitored optical parametric oscillator interacting with a non-zero temperature environment, and continuously monitored respectively via homodyne detection with $\vartheta_{\sf m}=0$ (red dotted line), homodyne detection with $\vartheta_{\sf m}=\pi/2$ (green solid line) and heterodyne detection (blue dashed line). Parameters are fixed as follows: 
$\chi/\kappa=0.4$, $n_{th}=1$, $n_0=2$.
}
\label{f:daemonicOPA_FiniteTemp}
\end{figure}
We observe that as expected the three strategies yield the same amount of steady-state daemonic ergotropy, as eventually all unravellings purify the initial state; however, during the dynamics, the results are much different. If we restrict to homodyne unravellings, our 
numerics show that a hierarchy is immediately apparent:  the optimal homodyne detection is obtained with phase $\vartheta_{\sf m}=\pi/2$, while the worst results are obtained by fixing $\vartheta_{\sf m}=0$. Remarkably, we have thus demonstrated that homodyne strategies leading to larger daemonic ergotropy (and thus larger purity) are the ones leading to lower amount of squeezing. The interpretation of the result is the following: we first remind ourselves that for states with zero correlations between $\hat{x}$ and $\hat{p}$ (that is with diagonal CMs), the purity of the state is inversely proportional to the product between the quadrature uncertainties $(\langle \Delta \hat{x}^2\rangle\langle\Delta\hat{p}^2\rangle)^{1/2}$. Purifying thus corresponds to jointly decrease the fluctuations of the two quadratures. If the Hamiltonian dynamics is already allowing to reduce the variance along the $\hat{x}$ quadrature, our results show that in order to faster purifying the state one should exploit the measurement to reduce the fluctuation of the orthogonal quadrature, rather than exploiting the measurement to further increase the squeezing of $\hat{x}$. On the other hand heterodyne detection, being phase invariant, simultaneously reduces the fluctuations for both quadratures, and one can expect a different and non-trivial behaviour respect to the homodyne strategies. For the values of parameters we have considered, we observe how the heterodyne strategy yields, in fact, values of daemonic ergotropy similar to the optimal homodyne strategy; as can be observed in the inset,  heterodyne is slightly better at small times, with homodyne becoming the optimal unravelling for $\kappa t \gtrsim 0.96$.

Let us now focus on the scenario where the environment temperature is larger than zero ($n_{th}=1$), whose numerical results are shown in Fig.~\ref{f:daemonicOPA_FiniteTemp}. As before, if we restrict to homodyne unravellings we still observe a well-defined hierarchy, showing that larger daemonic ergotropy (and thus larger purity) are obtained for homodyne phase $\vartheta_{\sf m} = \pi/2$, that is for the unravelling yielding lower squeezing; as expected, at steady-state both homodyne strategies yield the same amount of daemonic ergotropy, quantified by Eq.~\eqref{eq:daemonicOPA_Homo}. However, we also find that for these values of the parameters heterodyne is clearly the optimal strategy, not only at steady-state, but also during the whole dynamics. 
\section{Conclusions and outlook}\label{s:conclusions}

We have seen that the task of maximising the average conditional (i.e., `daemonic') ergotropy is, for single-mode bosonic Gaussian states, equivalent to the maximisation of the conditional purity of the state. Given that all entropies of one-mode Gaussian states are a function of their single symplectic eigenvalue and thus induce the same hierarchy on the set of states, this maximisation amounts to optimising the parametric cooling (purification) of the state through general-dyne measurements. And so the present study, by providing one with solutions for the most iconic and common Gaussian scenarios, determines both the maximum  extractable energy and the optimal general-dyne cooling scheme in all such cases. 

The multimode scenario of our problem is more delicate, as the general-dyne minimisation of the sum of the symplectic eigenvalues that has to be determined is far from trivial for multiple degrees of freedom. Notice also that, although such a sum does retain a clear entropic quality, being a sum of thermal quadrature noises, it does not technically correspond to any -- von Neumann, linear or Rényi -- entropy. The conditional minimisation of such a sum would hence be a very interesting endeavour.

Another scenario that could be the worthwhile focus of future investigation is the further enhancement of the ergotropy of continuously monitored systems through adaptive, rather than fixed, general-dyne strategies, which have not been considered here.
\section{Acknowledgements}
MGG would like to thank F. Albarelli for useful discussions and acknowledges support from MUR and Next Generation EU via the PRIN 2022 Project CONTRABASS (Contract N.2022KB2JJM) and NQSTI-Spoke2-BaC project QMORE (contract n. PE00000023-QMORE).

\bibliography{apssamp}

\end{document}